\documentclass[10pt]{article}

\usepackage[T2A]{fontenc}
\usepackage[english,russian]{babel}
\usepackage[utf8]{inputenc}

\usepackage{amsfonts}
\usepackage{amssymb}
\usepackage{mathrsfs}
\usepackage{amsmath}
\usepackage{tikz}\usetikzlibrary{matrix,fit,positioning, shapes, backgrounds}
\usepackage{setspace}
\usetikzlibrary{decorations.markings}

\usepackage{indentfirst}

\usepackage{dsfont}
\usepackage{esint}
\usepackage{hyperref}
\hypersetup{
    colorlinks = true,
    linkcolor = [rgb]{0 .4 .8},
    citecolor=[rgb]{1 .2 .2}
}
\usepackage{array}

\usepackage[numbers,sort]{natbib}
\setcitestyle{square}

\newcommand{\bea}{\begin{equation}}
\newcommand{\eea}{\end{equation}}
\newcommand{\bear}{\begin{eqnarray}}
\newcommand{\eear}{\end{eqnarray}}
\newcommand{\bearr}{\begin{eqnarray*}}
\newcommand{\eearr}{\end{eqnarray*}}

\newcommand{\beal}{\begin{align}}
\newcommand{\eeal}{\end{align}}
\newcommand{\beall}{\begin{align*}}
\newcommand{\eeall}{\end{align*}}

\newcommand{\CP}{\mathds{C}\mathds{P}}
\newcommand{\CC}{\mathds{C}}

\newcommand{\dd}{\partial}

\newcommand{\comment}[1]{}

\usepackage{xcolor}
\usepackage{empheq}

\newcolumntype{L}[1]{>{\raggedright\let\newline\\\arraybackslash\hspace{0pt}}m{#1}}
\newcolumntype{C}[1]{>{\centering\let\newline\\\arraybackslash\hspace{0pt}}m{#1}}
\newcolumntype{R}[1]{>{\raggedleft\let\newline\\\arraybackslash\hspace{0pt}}m{#1}}

\makeatletter
\def\@seccntformat#1{\@ifundefined{#1@cntformat}%
{\csname the#1\endcsname\quad}
{\csname #1@cntformat\endcsname}
}
\def\section@cntformat{{\normalfont\Large\thesection.}\quad}
\def\subsection@cntformat{\textsection\, \thesubsection.\quad}
\def\subsubsection@cntformat{\textsection\textsection\, \thesubsubsection.\quad}
\makeatother

\newsavebox\MBox

\newcommand*{\TitleFont}{%
      \usefont{\encodingdefault}{\rmdefault}{b}{n}%
      \fontsize{14}{20}%
      \selectfont}
 
\usepackage{sectsty}
 
\usepackage{bm}

\DeclareBoldMathCommand\boldlangle{\left\langle}
\DeclareBoldMathCommand\boldrangle{\right\rangle}

\sectionfont{\fontsize{12}{15}\selectfont}

\begin{document}
\title{\vspace{-1.5cm}\TitleFont Flag manifold sigma-models and nilpotent orbits\footnote{Invited contribution to the special issue of the ``Proceedings of the Steklov Institute of Mathematics'' dedicated to the 80-th anniversary of A.~A.~Slavnov.}}
\author{Dmitri Bykov\footnote{Emails:
bykov@mpp.mpg.de, bykov@mi-ras.ru}  \\ \\
{\small $\bullet$ Max-Planck-Institut f\"ur Physik, F\"ohringer Ring 6
D-80805 M\"unchen, Germany} \\ {\small $\bullet$ Steklov Mathematical Institute of Russ.~Acad.~Sci.,} \\{\small Gubkina str. 8, 119991 Moscow, Russia \;}}
\date{}
\maketitle

\begin{center}
\emph{To Andrei Alekseevich Slavnov on the occasion of his 80-th birthday, \\with respect and gratitude}
\end{center}

\maketitle
\vspace{-1cm}
\begin{center}
\line(1,0){370}
\end{center}
\vspace{-0.2cm}
\textbf{Abstract.} In the present paper we study flag manifold sigma-models that admit a zero-curvature representation. It is shown that these models may be naturally considered as interacting (holomorphic and anti-holomorphic) $\beta\gamma$-systems. Besides, using the theory of nilpotent orbits of complex Lie groups, we establish a relation to the principal chiral model.

\vspace{-0.7cm}
\begin{center}
\line(1,0){370}
\end{center}

\section{The gauged linear formulation and the  $\beta\gamma$-systems}

In the present paper we will consider sigma-models, whose target spaces are flag manifolds -- homogeneous manifolds of the group $SU(N)$ of the form
\bea\label{flagquot}
\mathcal{F}(n_1, ... ,n_m):=\frac{SU(N)}{S(U(n_1)\times \ldots \times U(n_m))},\quad\quad \sum\limits_{i=1}^m\,n_i=N\,.
\eea
Sometimes for brevity we will also denote this quotient as $G\over H$. This space admits a reductive metric  -- the so-called normal metric, which will be denoted by $\mathscr{G}$. It is defined as follows. Let $\mathfrak{g}=\mathfrak{h}\oplus \mathfrak{m}$ be the standard decomposition of the Lie algebra ($\langle \mathfrak{m}, \mathfrak{h}\rangle=0$, where $\langle . , .\rangle$ is the Killing metric). Then the normal metric on $G\over H$ is induced from the restriction $\langle . , .\rangle\big|_{\mathfrak{m}}$ of the Killing metric. An interesting fact is that in this metric all geodesics are homogeneous, i.e. they are orbits of one-parametric subgroups of  $G$~\cite{Alekseevsky}. Quite generally, integrability of the geodesic flow on $G\over H$ is a necessary condition for the integrability of the sigma-model.

In order to define the models we will also require a $G$-invariant (homogeneous) complex structure $\mathscr{I}$ on the target space. On the flag manifold~(\ref{flagquot}) such complex structures are in one-to-one correspondence with the orderings of the set $(n_1, \ldots , n_m)$. Let us assume that the ordering is given by the index $i$ of $n_i$, otherwise we simply swap the indices. Then the space~(\ref{flagquot}) allows an alternative definition as the quotient
\bea\label{complexquot}
\mathcal{F}(n_1, ... ,n_m)\underset{\mathscr{I}}{\simeq} {GL(N, \CC)\over P_{d_1, \ldots , d_m}}\,,
\eea
where $P_{d_1, \ldots , d_m}$ is a parabolic subgroup of $GL(N, \CC)$ that stabilizes the flag of linear spaces $0\subset L_1 \subset \ldots \subset L_m\simeq \CC^N$, $L_k \simeq \CC^{d_k}$ and $d_k=\sum\limits_{i=1}^k\,n_i$.

The action schematically has the following form~\cite{BykovCompl} (here $X:\Sigma\to \mathscr{F}$):
\bea\label{action}
\mathcal{S}[\mathscr{G}, \mathscr{I}]:=\int_\Sigma\,d^2 z\,\|\dd X\|^2_\mathscr{G}+\int_\Sigma\,X^\ast \omega,
\eea
where $\omega=\mathscr{G}\circ \mathscr{I}$ is the fundamental Hermitian form of the metric $\mathscr{G}$. It is closed if and only if $m=2$, i.e. when the flag manifold is a Grassmannian~\cite{GLSM1}. In other cases the second term in the action~(\ref{action}) is not topological.

The $B$-field of the same form as in~(\ref{action}) was considered on other grounds in~\cite{WittenTop}. Besides, Lax pairs for models of type~(\ref{action}), in the special case of symmetric spaces, were seemingly considered for the first time in~\cite{Bytsko} (the $\CP^1$-case) and in~\cite{BroZag} (mostly in the non-compact case). A detailed study of a similar case of para-compact target spaces may be found in the recent work~\cite{Delduc}.

In~\cite{GLSM1, GLSM2, BykovAnomaly} we constructed the gauged linear sigma-model representation for the models of type~(\ref{flagquot})--(\ref{action}). In the case when the target space is a Grassmannian, the metric $\mathscr{G}$ is K\"ahler, and this representation is equivalent to the K\"ahler quotient $G(k, N)\simeq \mathrm{Hom}(\CC^k, \CC^N)//U(k)$. In the general case our construction leads to a quotient w.r.t. a non-reductive group and to the ``Killing'' metric $\mathscr{G}$, which is not K\"ahler in general.

The construction is as follows. We introduce the field $U\in \mathrm{Hom}(\CC^M, \CC^N)$, where $M=d_{m-1}$, satisfying the orthonormality condition $U^\dagger U=\mathds{1}_M$, as well as the ``gauge'' field $\mathcal{A}=\mathcal{A}_z dz+\mathcal{A}_{\bar{z}} d\bar{z}$ of the following special form:
\bea\label{colormatr}
\mathcal{A}_z=
\begin{tikzpicture}[
baseline=-\the\dimexpr\fontdimen22\textfont2\relax,scale=1]
\matrix[matrix of math nodes,left delimiter=(,right delimiter=),ampersand replacement=\&] (matr) {
\node (X0) {\ast};\&\ast\&\ast\&\ast\&\ast\&\node (X00) {\ast};\\
\node (X1) {\ast};\& \node (X2) {\ast};\&\ast\&\ast\&\ast\&\node (X11) {\ast};\\
\&\&\ast\&\ast\&\ast\&\ast\\
\&\&\node (X3) {\ast};\&\node (X4) {\ast};\&\ast\&\node (X22) {\ast};\\
\&\&\&\&\ast\&\ast\\
\&\&\&\&\node (X5) {\ast};\&\node (X6) {\ast};\\
};
\node[fill=blue!10, fit=(X0.north west) (X11.south east) ] {};
\node[fill=blue!10, fit=(X2.south east) (X22.south east) ] {};
\node[fill=blue!10, fit=(X4.south east) (X6.south east) ] {};
\draw[blue!50, line width=1pt, rounded corners] ([yshift=-3pt,xshift=-3pt]X1.south west)  -- ([yshift=-3pt,xshift=-2.5pt]X2.south east) node[right,black] {};
\draw[blue!50, line width=1pt, rounded corners] ([yshift=-3pt,xshift=-3pt]X2.south east)  -- ([yshift=-3pt,xshift=-3pt]X3.south west) node[right,black] {};
\draw[blue!50, line width=1pt, rounded corners] ([yshift=-3pt,xshift=-3pt]X3.south west)  -- ([yshift=-3pt,xshift=-2.5pt]X4.south east) node[right,black] {};
\draw[blue!50, line width=1pt, rounded corners] ([yshift=-3pt,xshift=-3pt]X4.south east)  -- ([yshift=-3pt,xshift=-3pt]X5.south west) node[right,black] {};
\draw[blue!50, dotted, line width=1pt] ([yshift=-3pt,xshift=-3pt]X3.south west)  -- ([yshift=-40pt,xshift=-3pt]X3.south west) node[right,black] {};
\draw[blue!50, dotted, line width=1pt] ([yshift=-3pt,xshift=-3pt]X5.south west)  -- ([yshift=-10pt,xshift=-3pt]X5.south west) node[right,black] {};
\draw[latex-latex] ([yshift=-6pt,xshift=-3pt]X5.south west) --  node[below, yshift=2pt] {{\scriptsize$ d_1$}} ([yshift=-6pt,xshift=3pt]X6.south east);
\draw[latex-latex] ([yshift=-37pt,xshift=-3pt]X3.south west) --  node[below, yshift=2pt] {{\scriptsize$ d_2$}} ([yshift=-37pt,xshift=50pt]X3.south west);
\draw[latex-latex] ([yshift=6pt,xshift=-3pt]X0.north west) --  node[above, yshift=-2pt] {{\scriptsize$ d_{m-1}$}} ([yshift=6pt,xshift=3pt]X00.north east);
\end{tikzpicture},\quad\quad\quad \mathcal{A}_{\bar{z}}=(\mathcal{A}_z)^\dagger\,.
\eea
The Lagrangian reads
\bea\label{lagr2}
\mathscr{L}=\mathrm{Tr}\left(\|D_{\bar{z}} U\|^2\right)\,,\quad\quad\textrm{where}\quad\quad D_{\bar{z}} U=\dd_{\bar{z}}U+i\, U\mathcal{A}_{\bar{z}}\,.
\eea
This Lagrangian is equivalent to~(\ref{action}) -- it can be proven by eliminating the field $\mathcal{A}$. Due to the orthonormality condition $U^\dagger U=\mathds{1}_M$ the gauge group of the model is $U(n_1)\times \cdots \times U(n_m)$. A natural question is whether one can instead use a quotient w.r.t. the complex group of upper/lower-block-triangular matrices. To answer this, we give up the orthonormality condition and assume that $U\in \mathrm{Hom}(\CC^M, \CC^N)$ is an arbitrary complex matrix of rank $M$. We then write down the following Lagrangian:
\bea\label{lagr3}
\mathscr{L}=\mathrm{Tr}\left((D_{\bar{z}} U)^\dagger\,D_{\bar{z}} U\,\frac{1}{U^\dagger U}\right)\,.
\eea
It is easy to see that it is invariant w.r.t. complex gauge transformations  $U\to U g$, where  $g\in P_{d_1, \ldots, d_{m-1}}$. The Gram-Schmidt orthogonalization procedure brings the Lagrangian~(\ref{lagr3}) to the form~(\ref{lagr2}), but for a number of reasons the complex form is preferable. In order to get rid of the denominator in the Lagrangian, we introduce an auxiliary field $V\in \mathrm{Hom}(\CC^N, \CC^M)$ and write down a new Lagrangian
\bea\label{lagr4}
\mathscr{L}=\mathrm{Tr}\left(V D_{\bar{z}} U\right)+\mathrm{Tr}\left(V D_{\bar{z}} U\right)^\dagger-\mathrm{Tr}\left(VV^\dagger U^\dagger U\right)\,,
\eea
that turns into the original one upon elimination of the field $V$. Next we perform yet another quadratic transformation, in order to eliminate the quartic interaction. To this end we introduce the complex matrix field $\Phi_z\in \mathrm{End}(\CC^N)$ and its Hermitian conjugate: $\Phi_{\bar{z}}=(\Phi_z)^\dagger$. We write one more Lagrangian
\bea\label{lagr5}
\mathscr{L}=\mathrm{Tr}\left(V \mathscr{D}_{\bar{z}} U\right)+\mathrm{Tr}\left(V \mathscr{D}_{\bar{z}} U\right)^\dagger+\mathrm{Tr}\left(\Phi_z \Phi_{\bar{z}}\right)\,,
\eea
where $\mathscr{D}_{\bar{z}}$ is the ``elongated'' covariant derivative
\bea
\mathscr{D}_{\bar{z}} U=\dd_{\bar{z}} U+i \,U \mathcal{A}_{\bar{z}}+i\,\Phi_{\bar{z}} U\,.
\eea
Let us clarify the geometric meaning of the Lagrangian~(\ref{lagr5}). The first two terms correspond to a sum of the so-called $\beta\gamma$-systems on the flag manifold $\mathcal{F}$, in a background field $\Phi_z$~\cite{Nekrasov, WittenBeta}. By definition, such a system may be defined for an arbitrary complex manifold $\mathcal{M}$ ($\mathrm{dim}_\CC \mathcal{M}=m$) with the help of a complex fundamental $(1,0)$-form $\theta=\sum\limits_{i=1}^m\,p_i\,dq_i$ on $T^\ast \mathcal{M}$. Here $q_i$ are the complex coordinates on $\mathcal{M}$ and  $p_i$ are the complex coordinates in the fiber of the  holomorphic cotangent bundle. The action of the $\beta\gamma$-system is then simply $S=\int\limits_\Sigma\,d^2z\,\sum\limits_{i=1}^m\,p_i\,\dd_{\bar{z}}q_i$. In the case of the flag manifold this action can be most conveniently written, using two matrices $U\in \mathrm{Hom}(\CC^M, \CC^N), V\in \mathrm{Hom}(\CC^N, \CC^M)$ and the gauge field $\mathcal{A}_z$. Indeed, it will be shown in the next section that the fundamental $(1,0)$-form can be written as $\theta=\mathrm{Tr}\left(V dU\right)\big|_{\mu_{\CC}=0}$, where $(U, V)$ satisfy the condition
\bea\label{mommapzero}
\mu_{\CC}=VU\big|_{\mathfrak{k}^\ast}=0\,, \quad\quad\textrm{and}\quad\quad \mathfrak{k}=\mathrm{Lie}(P_{d_1, \ldots, d_{m-1}})
\eea
is the Lie algebra of the corresponding parabolic subgroup of $GL(M, \CC)$. It is also assumed that the space of matrices, satisfying this condition, is factorized w.r.t. the action of $P_{d_1, \ldots, d_{m-1}}$, i.e. one has a complex symplectic reduction. The condition~(\ref{mommapzero}) is precisely the condition of vanishing of the moment map $\mu_\CC=0$ for the action of the parabolic group $P_{d_1, \ldots, d_{m-1}}$ on the space of matrices $(U, V)$ endowed with the symplectic form $\omega_0=\mathrm{Tr}(dU\wedge dV)$. As a result,
\bea\label{omegared}
d\theta=\omega_{\mathrm{red}}
\eea
is the complex symplectic form, arising after the reduction w.r.t. the parabolic group. In order to ensure the condition~(\ref{mommapzero}) at the level of the Lagrangian of the model, one needs the gauge field $\mathcal{A}_{\bar{z}}\in \mathrm{Lie}(P_{d_1, \ldots, d_{m-1}})$. Indeed, differentiating the Lagrangian~(\ref{lagr5}) w.r.t. $\mathcal{A}_{\bar{z}}$, one arrives at the condition~(\ref{mommapzero}).

A large class of integrable sigma-models in the $\beta\gamma$-formulation has been recently proposed in~\cite{Costello}~(for background material see also~\cite{Costello1, Costello2}). In the terminology of that work our field $\Phi_{\bar{z}}$ should be viewed as the component $A_{\bar{w}}$ of the Chern-Simons gauge field along the ``topological plane'' (i.e. the worldsheet~$\Sigma$). The quadratic form in the interaction term  $\mathrm{Tr}\left(\Phi_z \Phi_{\bar{z}}\right)$ in~(\ref{lagr5}) is in this context the inverse propagator of the field $A_{\bar{w}}$, which in the present (rational) case is proportional to the identity matrix.

\subsection{Relation to the quiver formulation}

Before passing to further topics, let us clarify the relation of the complex symplectic form, constructed using the symplectic quotient w.r.t. a parabolic subgroup, as above, and the symplectic form that arises as a result of a reductive quotient, defined by the so-called quiver.  We recall that $T^\ast \mathscr{F}$ is a hyper-K\"ahler manifold that maybe be constructed by a hyper-K\"ahler quotient of flat space (though we stress that the real symplectic form -- the K\"ahler form -- will not concern us here). This quotient is based on a linear quiver diagram of the following form:
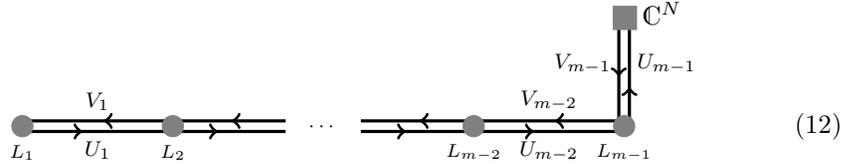
\bea
\begin{tikzpicture}[
baseline=-\the\dimexpr\fontdimen22\textfont2\relax,scale=1]
\begin{scope}[very thick,decoration={
    markings,
    mark=at position 0.4 with {\arrow{>}}}
    ] 
\draw[postaction={decorate}] ([yshift=-2pt,xshift=0pt]0,0) --  node [below, yshift=0pt] {\footnotesize $U_1$} ([yshift=-2pt,xshift=0pt]2,0);
\draw[postaction={decorate}] ([yshift=-2pt,xshift=0pt]2,0) --    ([yshift=-2pt,xshift=0pt]3.5,0);
\draw[postaction={decorate}] ([yshift=-2pt,xshift=0pt]4.5,0) --    ([yshift=-2pt,xshift=0pt]6,0);
\draw[postaction={decorate}] ([yshift=-2pt,xshift=0pt]6,0) --  node[below, yshift=0pt] {{\footnotesize$ U_{m-2}$}} ([yshift=-2pt,xshift=0pt]8,0);
\end{scope}
\begin{scope}[very thick,decoration={
    markings,
    mark=at position 0.6 with {\arrow{<}}}
    ] 
\draw[postaction={decorate}] ([yshift=2pt,xshift=0pt]0,0) --  node[above, yshift=0pt] {{\footnotesize $ V_1$}} ([yshift=2pt,xshift=0pt]2,0);
\draw[postaction={decorate}] ([yshift=2pt,xshift=0pt]2,0) --  ([yshift=2pt,xshift=0pt]3.5,0);
\draw[postaction={decorate}] ([yshift=2pt,xshift=0pt]4.5,0) --  ([yshift=2pt,xshift=0pt]6,0);
\draw[postaction={decorate}] ([yshift=2pt,xshift=0pt]6,0) --  node[above, yshift=0pt] {{\footnotesize$ V_{m-2}$}} ([yshift=2pt,xshift=0pt]8,0);
\end{scope}
\begin{scope}[very thick,decoration={
    markings,
    mark=at position 0.6 with {\arrow{<}}}
    ] 
\draw[postaction={decorate}] ([yshift=0pt,xshift=-2pt]8,0) --  node[right, xshift=3pt, yshift=5pt] {{\footnotesize$ U_{m-1}$}} ([yshift=0pt,xshift=-2pt]8,1.3);
\end{scope}
\begin{scope}[very thick,decoration={
    markings,
    mark=at position 0.4 with {\arrow{>}}}
    ] 
\draw[postaction={decorate}] ([yshift=0pt,xshift=2pt]8,0) --  node[left, xshift=-3pt, yshift=5pt] {{\footnotesize$ V_{m-1}$}} ([yshift=0pt,xshift=2pt]8,1.3);
\filldraw[black!50] (7.87,1.3) rectangle ++(8pt,8pt);
\end{scope}
\filldraw[black!50] (0,0) circle (4pt); \filldraw[black!50] (2,0) circle (4pt); \filldraw[black!50] (6,0) circle (4pt); \filldraw[black!50] (8,0) circle (4pt);
\node at (8.5,1.5) {$\CC^N$};
\node at (0,-0.35) {\scriptsize $L_1$};
\node at (2,-0.35) {\scriptsize $L_2$};
\node at (6,-0.35) {\scriptsize $L_{m-2}$};
\node at (8,-0.35) {\scriptsize $L_{m-1}$};
\node at (4,0) {\scriptsize $\cdots$};

\end{tikzpicture}
\eea

In each node there is a vector space $L_k\simeq \CC^{d_k}$, and to each arrow from node $i$ to node $j$ corresponds a field, taking values in $\mathrm{Hom}(L_i, L_j)$. The full space of fields is therefore
\bea
\mathscr{W}_0:=\oplus_{i=1}^{m-1}\,\left(\mathrm{Hom}(L_i, L_{i+1})\oplus \mathrm{Hom}(L_{i+1}, L_i)\right)\,.
\eea
In each node there is an action of a gauge group $GL(L_i)$. We then consider the GIT-quotient $\mathscr{W}_f:=\mathscr{W}/\mathscr{G}$ of the stable subset $\mathscr{W}\subset \mathscr{W}_0$ w.r.t. the group $\mathrm{G}:=\prod\limits_{i=1}^{m-1}\,GL(L_i)$. In $\mathscr{W}_f$ we define a submanifold given by the vanishing conditions for the moment maps ($U_0=0, V_0=0$):
\bea\label{mommapi}
\mathscr{F}:= \{\mu_i=U_{i-1}V_{i-1}-V_iU_i=0\,,\quad\quad i=1, \ldots, m-1\,\}\subset \mathscr{W}_f\,.
\eea
The (well-known) statement is that the resulting space is the flag manifold~(\ref{complexquot}), which is why we have denoted it by $\mathscr{F}$. On $\mathscr{W}_0$ there is a natural complex {symplectic~form}
\bea
\Omega=\sum\limits_{i=1}^{m-1}\,\mathrm{Tr}(dU_i \wedge dV_i)\,.
\eea
The construction just described may be interpreted as the symplectic quotient w.r.t. the complex group $\mathrm{G}$, and it endows $\mathscr{F}$ with a certain symplectic form $\Omega_\mathscr{F}$. We prove the following statement:

\vspace{0.3cm}
\noindent\textbf{Lemma.} $\Omega_\mathscr{F}=\omega_{\mathrm{red}}$, where $\omega_{\mathrm{red}}$ is the symplectic form~(\ref{omegared}) that arises as a result of the reduction w.r.t. a parabolic subgroup of $GL(M, \CC)$.

\vspace{0.3cm}
\noindent\textbf{Proof.} First of all, consider the fields $\{U_i\}$. $U_i$ is a matrix with $d_i$ columns and $d_{i+1}$ rows. By the action of $GL(d_{i+1}, \CC)$ one can bring $U_i$ to the form where the first $d_{i+1}-d_i$ rows are zero and the last $d_i$ rows represent a unit matrix. The stabilizer of this canonical form w.r.t. the joint (left-right) action of $GL(d_{i+1}, \CC)\times GL(d_i, \CC)$ is the subgroup $P_{d_i, d_{i+1}} \subset P_{d_i, d_{i+1}}\times GL(d_{i}, \CC)$, embedded according to the rule $g\to (g, \pi_i(g))$, where $\pi_i(g)$ is the projection on the block of size $d_i\times d_i$. Iterating this procedure, i.e. bringing all matrices  $U_i$ ($i=1, \ldots, m-2$) to canonical form, we arrive at the situation, when one is left with a single non-trivial matrix $U_{m-1}:=U$, and the resulting symmetry group is precisely $P_{d_1, \ldots, d_{m-1}}$. We also denote $V_{m-1}:=V$. Now, let $a\in \mathfrak{k}=\mathrm{Lie}(P_{d_1, \ldots, d_{m-1}})$. By definition of the stabilizer $a U_{m-2}=U_{m-2}\pi_{m-2}(a)$, therefore $\mathrm{Tr}(aU_{m-2}V_{m-2})=\mathrm{Tr}(\pi_{m-2}(a)V_{m-2}U_{m-2})=\mathrm{Tr}(\pi_{m-2}(a)U_{m-3}V_{m-3})$, where in the second equality we have used the equation~(\ref{mommapi}). Since $\pi_{m-2}(a)\in \mathrm{Stab}(U_{m-3})$, we can iterate this procedure, and at the end we will obtain $\mathrm{Tr}(aU_{m-2}V_{m-2})=0$. Due to the equation $U_{m-2}V_{m-2}-VU=0$ we get $VU\big|_{\mathfrak{k}^\ast}=0$, which coincides with~(\ref{mommapzero}). Besides, since the matrices $U_i$ ($i=1, \ldots, m-2$) are constant, the restriction of the symplectic form $\Omega$ coincides with $\mathrm{Tr}(dU_{m-1}\wedge dV_{m-1})=\mathrm{Tr}(dU\wedge dV)$. \dotfill $\blacksquare$

\vspace{0.5cm}
Let us clarify the role of the field $\Phi_z$. Differentiating the Lagrangian~(\ref{lagr5}) w.r.t. $\Phi_{\bar{z}}$, we obtain
\bea\label{Phiz}
\Phi_z=-i\,UV\,.
\eea
This coincides with the expression for the $z$-component of the Noether current for the action of the group  $GL(N, \CC)$ on the space of matrices $(U, V)$. For the $\beta\gamma$-system written above $\Phi_z$ is nothing but the moment map for the action of this group.

\section{Relation to the principal chiral model}

Recall that the equations of motion of the principal chiral model may be written as follows:
\bea
j=-g^{-1}dg,\quad\quad d\ast j=0\,.
\eea
Analogously one can write down the equations of motion for a sigma-model with a symmetric target space. Let $\sigma: G\to G$ be the Cartan homomorphism, $\sigma^2=1$. Then the equations of motion have the form:
\bear
j_s=-\widehat{g}^{-1}d\widehat{g},\quad\quad d\ast j_s=0\,,\\
\textrm{where}\quad\quad \widehat{g}=\sigma(g)g^{-1}.
\eear
Note that the map $g\to \widehat{g}$, written in the second line, is nothing but the Cartan embedding $\hat{\sigma}: {G\over H} \hookrightarrow G$. In both cases $j, j_s$ are the Noether currents of the sigma-model, calculated using the standard action $S=\int \,d^2z\,\|\dd X\|^2$.

In other words, suppose $X: \Sigma \to G$ is a harmonic map. If its image lies in the symmetric space $G\over H$, i.e. $X(\Sigma)\subset {G\over H}\underset{\hat{\sigma}}{\subset} G$, the map $X: \Sigma \to {G\over H}$ is harmonic. The converse is also true: if $X: \Sigma \to {G\over H}$ is a harmonic map, $\hat{\sigma} \circ X: \Sigma \to G$ is also harmonic. This can be alternatively understood by recalling that $\hat{\sigma}$ is a totally geodesic embedding. By definition, this means that the second fundamental form of $\hat{\sigma}({G\over H})\subset G$ vanishes: $(\nabla_X Y)^\perp=0$ for any two vectors $X, Y\in T(\hat{\sigma}({G\over H}))$. It is easy to check that if $\hat{\sigma}: \mathcal{M}\subset \mathcal{N}$ is a totally geodesic submanifold, and $X: \Sigma \to \mathcal{M}$ is a harmonic map, then $\hat{\sigma} \circ X: \Sigma \to \mathcal{N}$ is also harmonic.

\subsection{Nilpotent orbits}

\noindent
After this intermezzo let us return to the formulas~(\ref{mommapzero})-(\ref{Phiz}):
\bea\label{PhizUV}
\Phi_z=-i\,UV\,,\quad\quad \mu_{\CC}=VU\big|_{\mathfrak{k}^\ast}=0\,,
\eea
where $\mathfrak{k}=\mathrm{Lie}(P_{d_1, \ldots, d_{m-1}})$.

\subsubsection{Grassmannian}
To start with, we consider the case of a Grassmannian, i.e. $m=2$. Then the vanishing of the moment map is simply $VU=0$. Therefore $\Phi_z^2=0$. From the expression for $\Phi_z$ it also follows that $\mathrm{Im}(\Phi_z)\subset\mathrm{Im}(U)\subset \mathrm{Ker}(\Phi_z)$. As is well-known,
\bea\label{cotgrass}
\{(U, \Phi_z): \mathrm{rk}(U)=M, \;\Phi_z^2=0, \;\;\mathrm{Im}(\Phi_z)\subset\mathrm{Im}(U)\subset \mathrm{Ker}(\Phi_z)\} \simeq T^\ast G(M, N)
\eea
is the cotangent bundle to a Grassmannian, and the forgetful map 
\bea
T^\ast G(M, N)\to \{\Phi_z: \Phi_z^2=0\}
\eea
provides a resolution of singularities of the nilpotent orbit in the r.h.s. (the Springer resolution). The conditions in the l.h.s. of~(\ref{cotgrass}) imply the factorization~(\ref{PhizUV}) for $\Phi_z$, and the non-uniqueness in this factorization corresponds exactly to the gauge symmetry $U\to U g, V\to g^{-1} V$, where $g\in GL(M, \CC)$.

Let us now derive the equations of motion for the field $\Phi_z$. First of all, the Lagrangian~(\ref{lagr5}) implies the following equations of motion for the fields $U$ and $V$:
\bear
\mathscr{D}_{\bar{z}} U=0,\quad\quad \mathscr{D}_{\bar{z}} V=0\,.
\eear
Therefore $\mathscr{D}_{\bar{z}} \Phi_z=0$, i.e.
\bea\label{dPhi}
\dd_{\bar{z}}\Phi_z +i\,[\Phi_{\bar{z}}, \Phi_z]=0\,.
\eea
This equation is nothing but the equation of motion of the principal chiral field. Indeed, introduce a 1-form $j=i(\Phi_z\,dz+\Phi_{\bar{z}}\,d\bar{z})$ with values in the Lie algebra $\mathfrak{u}_N$. In this case~(\ref{dPhi}) together with the Hermitian conjugate equation may be written in the form of two conditions
\bea
d\ast j=0,\quad\quad dj-j\wedge j=0\,,
\eea
which are the e.o.m. of the principal chiral field. This is consistent with the fact, proven in~\cite{BykovCompl}, that the Noether current of the model~(\ref{action}) is flat.

In other words, in the case of a Grassmann sigma-model the field $\Phi_z$ satisfies the equations
\bea\label{nilpeqs}
\dd_{\bar{z}}\Phi_z +i\,[\Phi_{\bar{z}}, \Phi_z]=0\,,\quad\quad \Phi_z^2=0\,.
\eea
It is rather clear, though, that these equations do not completely characterize the Grassmannian sigma-model. Indeed, for example the solution $\Phi_z=0$ of the equations~(\ref{nilpeqs}) corresponds in fact to a whole large class of sigma-model solutions: $D_{\bar{z}} U=0,\; D_{\bar{z}} V=0,\; UV=0$. Since, according to the assumption, $U^\dagger U$ is a non-degenerate matrix, it follows from the latter condition that $V=0$, and as a result one is left with the equation $D_{\bar{z}} U=0$, i.e. a holomorphic map $U$ to $G(M, N)$. The converse is also true: if $D_{\bar{z}} U=0$, then $\Phi_{\bar{z}}U=0$, and it follows from $\Phi_z=UV$ that $\Phi_{\bar{z}} \Phi_z=\Phi_z^\dagger \Phi_z=0$, i.e. $\Phi_z=0$. As a result, one has the correspondence
\bea
\Phi_z=0\quad\quad \leftrightarrow\quad\quad \textrm{holomorphic curves in}\;\;G(M, N)\,.
\eea
Therefore the equation for $U$ is essential in general. As for the field $V$, it can be completely excluded and replaced by the field $\Phi_z$. Indeed, from the conditions in the l.h.s. of~(\ref{cotgrass}) one obtains the factorization $\Phi_z=UV$, and the equations $\mathscr{D}_{\bar{z}}U=0$, $\mathscr{D}_{\bar{z}}\Phi_z=0$ imply $U \mathscr{D}_{\bar{z}}V=0$. The condition that  $U$ is a matrix of rank $M$ leads to $\mathscr{D}_{\bar{z}}V=0$. In other words, an alternative formulation of the model is as follows: we consider the fields $(U, \Phi_z)$, satisfying
\bea\label{conds1}
(U, \Phi_z): \mathrm{rk}(U)=M, \;\Phi_z^2=0, \;\;\mathrm{Im}(\Phi_z)\subset\mathrm{Im}(U)\subset \mathrm{Ker}(\Phi_z)\,.
\eea
In this case the equations of motion take the form
\bea
\mathscr{D}_{\bar{z}} U=0,\quad\quad \mathscr{D}_{\bar{z}} \Phi_z=0\,.
\eea
Next we describe the situation when the equation for $U$ is indeed redundant:

\vspace{0.3cm}
\noindent\textbf{Lemma.} The equation $\mathscr{D}_{\bar{z}} U=0$ follows from $\mathscr{D}_{\bar{z}} \Phi_z=0$ and from the conditions~(\ref{conds1}) if and only if $\mathrm{Ker}(\Phi_z)\simeq \mathrm{Im}(U)$.

\vspace{0.3cm}
\noindent\textbf{Proof.} Multiplying $D_{\bar{z}} U$ by $\Phi_z$ and using the equation $D_{\bar{z}} \Phi_z=0$ and the condition $\Phi_z U=0$, we obtain $\Phi_z D_{\bar{z}} U=D_{\bar{z}} (\Phi_z U)=0$. Therefore the equation $\Phi_zD_{\bar{z}} U=0$ holds identically. It is equivalent to $\mathscr{D}_{\bar{z}} U=0$ if and only if $\mathrm{Ker}(\Phi_z)\simeq \mathrm{Im}(U)$. \dotfill $\blacksquare$

\vspace{0.3cm}
The condition $\Phi_z^2=0$ means that the Jordan structure of  $\Phi_z$ consists of $m$ cells of sizes $2\times 2$ and $n$ cells of sizes $1\times 1$. In this case $N=2m+n$ and $\mathrm{dim \;Ker}(\Phi_z)=m+n$. Since $\mathrm{Im}(U)\subset \mathrm{Ker}(\Phi_z)$ and $\mathrm{rk}(U)=M$, we get the condition $M\leq m+n$. This easily leads to $m\leq N-M$, $n\geq 2M-N$, and the inequalities are saturated precisely in the case $\mathrm{Ker}(\Phi_z)\simeq \mathrm{Im}(U)$ considered earlier. In this case the number of $2\times 2$ cells is maximal and equal to $N-M$, and the number of cells of size $1\times 1$ is $2M-N$. Note that this is only possible in the case $M\geq {N\over 2}$. Reduction of the number of cells of type $2\times 2$ corresponds to the degeneration of the matrix $\Phi_z$.

The dynamical equation~(\ref{nilpeqs}) imposes severe constraints on the way, in which the Jordan structure of the matrix  $\Phi_z$ can change as one varies the point $z, \bar{z}$ on the worldsheet. Indeed, it implies that $\Phi_z=kQ(z)k^{-1}$, where $Q(z)$ is a matrix that depends holomorphically on $z$. The Jordan structure of the matrix $Q(z)$ is the same as that of $\Phi_z$, and the vanishing of the Jordan blocks occurs holomorphically in~$z$. In particular, the Jordan structure changes only at ``special points'' -- isolated points on the worldsheet. As a result, ``almost everywhere'' the dimension of the kernel $\mathrm{dim\;Ker}(\Phi_z):=\widetilde{M}$ is the same, and the map $\lambda: (z, \bar{z})\to \mathrm{Ker}(\Phi_z)$ is a map to the Grassmannian $G(\widetilde{M}, N)$. A more careful analysis of the behavior of $\Phi_z$ at a special point would show that $\lambda$ may be extended to these points. We have come to the following conclusion: let $g(z, \bar{z})$ be a solution of the principal chiral model, i.e. a harmonic map to the group $G$, satisfying the condition $\Phi_z^2=0$, where $\Phi_z:=g^{-1}\dd_{z}g$ is a component of the Noether current, and let the dimension of the kernel of  $\Phi_z$ at a typical point of the worlsheet be $\widetilde{M}$. Then one can construct a harmonic map to the Grassmannian $G(\widetilde{M}, N)$ by the rule $(z, \bar{z}) \to \mathrm{Ker}(\Phi_z)$.

\subsubsection{The $SU(2)$-case.}
We consider the special case $N=2$, i.e. when $G=SU(2)$. In this case $\mathrm{Tr}(\Phi_z)=0$, and the condition $\Phi_z^2=0$ is equivalent to $\mathrm{Tr}(\Phi_z^2)=0$. The latter condition is, in turn, the Virasoro constraint, i.e. the condition that the harmonic map is minimal. As a result, to a minimal surface in $G=SU(2)$, satisfying\footnote{Note that the condition $\Phi_z=0$ implies $g(z, \bar{z})=g_0$, i.e. in this case the map is trivial.} $\Phi_z\not\equiv 0$, we can prescribe a harmonic map into $\CP^1$. Let us construct it explicitly. The Cartan embedding $\CP^1 \hookrightarrow SU(2)$ has the form $g=\mathds{1}_2-2 w\otimes \bar{w}$, where $w\in \CP^1$ ($\|w\|=1$). The Noether current is $\Phi_z=2(w\otimes D_z \bar{w}-D_z w\otimes \bar{w})$, where $D_z w=\dd_z w-(\bar{w}\circ \dd_z w) w$, and its square is $\Phi_z^2=-4(D_z w\otimes D_z \bar{w}+(D_z  \bar{w}\circ D_z w) w\otimes \bar{w})$. Multiplying the condition $\Phi_z^2=0$ by $w$ from the right and,  taking into account that $D_z \bar{w}\circ w=0$, we get $D_z  \bar{w}\circ D_z w=0$, therefore $D_z w\otimes D_z \bar{w}=0$. We see that the map $w(z, \bar{z})$ is either holomorphic or anti-holomorphic. If $D_z \bar{w}=0$, the null-vector $\chi$ of the matrix $\Phi_z$ ($\Phi_z \chi=0$) satisfies the constraint $\bar{w}\circ \chi=0$, i.e. $\chi$ is the antipodal point to $w$ on $\CP^1$. If $D_z w=0$, then, as is easy to verify, $D_{\bar{z}}w$ is a holomorphic map, and the null-vector $\chi$ satisfies the condition $D_z \bar{w}\circ \chi=0$. Therefore in both cases we are led to the conclusion that $\chi$ is an anti-holomorphic map, related to a holomorphic map ($w$ or $D_{\bar{z}}w$) by the antipodal involution.

\subsubsection{The partial flag manifold}

Let us extend the results of the previous section to more general flag manifolds. We return first to the equation for $\Phi_z$:
\bea\label{dPhi1}
\dd_{\bar{z}}\Phi_z +i\,[\Phi_{\bar{z}}, \Phi_z]=0\,,
\eea
but this time we assume that the matrix $\Phi_z$ satisfies, in a typical point $(z, \bar{z})\in \Sigma$, the condition
\bea\label{phinilp}
\Phi_z^m=0 \quad\quad \textrm{and}\quad\quad \Phi_z^{m-1}\neq0\;.
\eea
The matrix $\Phi_z$ naturally defines a flag
\bea\label{kerflag}
f:=\;\;\{0\subset\mathrm{Ker}(\Phi_z)\subset \mathrm{Ker}(\Phi_z^2)\subset \cdots \subset \mathrm{Ker}(\Phi_z^m)\simeq \CC^N\}
\eea

\vspace{0.3cm}
\noindent\textbf{Proposition.} Given a matrix $\Phi_z$ satisfying~(\ref{dPhi1})-(\ref{phinilp}), the map $(z, \bar{z}) \to f$ is a solution to the e.o.m. of the flag manifold sigma-model~(\ref{lagr5}).

\vspace{0.3cm}
\noindent\textbf{Proof.} Consider a matrix $U$ of the form $U=(U_{m-1}|\cdots|U_{1})$, where $U_i$ is a matrix, whose columns are the linearly independent vectors from $\mathrm{Ker}(\Phi_z^i)/\mathrm{Ker}(\Phi_z^{i-1})$. Let us relate the dimensions of these spaces to the dimensions of the Jordan cells of the matrix $\Phi_z$. To this end we bring $\Phi_z$ to the Jordan form
\bea
\Phi_z^{(0)}=\mathrm{Diag}\{J_{s_1}, \ldots, J_{s_\ell}\},\quad\quad \sum\limits_{j=1}^{\ell} s_j=N\,,
\eea
where $J_{s}$ is a Jordan cell of size $s\times s$. We have chosen the ordering $s_1\geq \ldots \geq s_\ell$, where, according to the supposition~(\ref{phinilp}), $s_1=m$.  We denote by $\kappa_i$ the number of Jordan cells of size at least $i$ ($\kappa_1=\ell$). The following two properties are obvious:
\begin{itemize}
\item $\kappa_{i+1}\leq \kappa_i$, i.e. $\kappa_1, \ldots, \kappa_m$ is a non-increasing sequence.
\item $\mathrm{dim \,Ker}(\Phi_z)=\kappa_1$, $\mathrm{dim \,Ker}(\Phi_z^2)=\kappa_1+\kappa_2$ etc., \\therefore $\mathrm{dim \,Ker}(\Phi_z^i)/\mathrm{Ker}(\Phi_z^{i-1})=\kappa_i$.
\end{itemize}
Therefore $U_i\in \mathrm{Hom}(\CC^{\kappa_i}, \CC^N)$ and $U\in \mathrm{Hom}(\CC^{M}, \CC^N)$, where $M=\sum\limits\limits_{i=1}^{m-1} \kappa_i$.

Since, by construction, $\mathrm{Im}(U)\simeq \mathrm{Ker}(\Phi_z^{m-1})$ and $\mathrm{Im}(\Phi_z)\subset \mathrm{Ker}(\Phi_z^{m-1})$, we have $\mathrm{Im}(\Phi_z)\subset \mathrm{Im}(U)$, i.e. there exists a matrix $V\in \mathrm{Hom}(\CC^N, \CC^M)$, such that $\Phi_z=-i\,UV$. Let us now derive the equations of motion for the matrices $U$ and~$V$. Since $\Phi_z^k U_k=0$ and $D_{\bar{z}}\Phi_z=0$, one has $\Phi_z^k D_{\bar{z}} U_k=0$. The columns of the matrix $(U_k|\cdots | U_1)$ span the kernel of $\Phi_z^k$, hence $D_{\bar{z}} U_k=-i\,\sum\limits_{j\leq k}\,U_j \mathcal{A}_{jk}$, where $\mathcal{A}_{jk}$ are matrices of relevant sizes. Out of the matrices $\mathcal{A}_{jk}$ ($1\leq j\leq k\leq m-1$) we form a single matrix $\mathcal{A}_{\bar{z}}$, which schematically looks as in~(\ref{colormatr}). Then, clearly, the following equation is satisfied:
\bea
\mathscr{D}_{\bar{z}} U=\dd_{\bar{z}} U+i \,U \mathcal{A}_{\bar{z}}+i\,\Phi_{\bar{z}} U=0\,.
\eea
As $\Phi_z=-i\,UV$, from the non-degeneracy of $U$ it follows that $\mathscr{D}_{\bar{z}} V=0$. Because $U_1, \ldots, U_k \subset \mathrm{Ker}(\Phi_z^k)$, in the matrix $\Phi_z^k U\sim U(VU)^k$ the last $\sum\limits_{i=1}^k \kappa_i$ columns vanish, therefore the matrix $(VU)^k$ is strictly-lower-triangular and has zeros on the first $k$ block diagonals (the main diagonal is counted as the first one). We denote by $\mathfrak{k}$ the parabolic subalgebra of $\mathfrak{gl}_M$ that stabilizes the subflag of~(\ref{kerflag}) with the last element omitted. We have proven that $VU\big|_{\mathfrak{k}^\ast}=0$. Therefore a solution $\Phi_z(z, \bar{z})$ of the system~(\ref{dPhi1})-(\ref{phinilp}) produces a solution $(U, V)$ to the equations of motion of the sigma-model with target space the flag manifold
\bear
&&\!\!\!\!\!\!\!\!\!\!\!\!\!\!\!{U(N)\over U(\kappa_1)\times \cdots \times U(\kappa_m)}, \quad\quad \textrm{where}\\
\nonumber &&\!\!\!\!\!\!\!\!\!\!\!\!\!\!\!\kappa_j=\mathrm{dim \,Ker}(\Phi_z^j)/\mathrm{Ker}(\Phi_z^{j-1}) \quad \quad \textrm{is a non-increasing sequence.}
\eear
The complex structure on the flag is uniquely determined by the structure of the complex flag~(\ref{kerflag}). \dotfill $\blacksquare$
 
 \section{Discussion}
 
In the present paper we described two principal results. First of all, we showed that the flag manifold introduced earlier by the author allow an alternative formulation as two coupled $\beta\gamma$-systems, interacting via an auxiliary field $\Phi_z$. This proves the equivalence of these models to the flag manifold models described in~\cite{Costello} (in the terminology of that paper the field $\Phi_z$ should be interpreted as the holomorphic component of the gauge field  $A_w$ along the ``topological plane'' that coincides with the worldsheet $\Sigma$ of the sigma-model. We believe that the gauged linear formulation of the $\beta\gamma$-system for flag manifold models introduced in the present paper will also prove useful for the investigation of their trigonometric ($\eta$) deformations (there is a vast literature on the subject: see, for example, \cite{Fateev1, Fateev2, Klimcik1, Klimcik2, Delduc2, Lukyanov, Tseytlin}).

Besides, we investigated the relation between the flag manifold sigma-models and the principal chiral model. It was shown that the solutions of the principal chiral model, which define a map into the nilpotent orbit of the corresponding complexified group, correspond to solutions of the flag manifold sigma-models (see~\cite{Crooks} for a nice review of the theory of nilpotent orbits). It seems likely that the full analysis of this correspondence will require the theory of the Springer resolution~(see, for example,~\cite{Ginzburg}) and will be a subject of further investigation.

\vspace{0.5cm}
\noindent\textbf{Acknowledgments.}\\ \\
I would like to express my sincere gratitude to A.~A.~Slavnov for his yearslong support and warm attitude. I wish Andrei Alekseevich health, cheerfulness and the pleasure of creative work.\\ \\
I would like to thank A.~Bourget, R.~Donagi, S.~Frolov, A.~Hanany, C.~Klim\v{c}\'{\i}k, T.~McLoughlin, V.~Pestun, E.~Sharpe, S.~Shatashvili, S.~Theisen, A.~Tseytlin and P.~Zinn-Justin for useful discussions, as well as D.~L\"ust for support.
I am grateful to the Institut des Hautes \'Etudes Scientifiques (Bures-sur-Yvette, France), where part of this work was done, and in particular to V.~Pestun, for hospitality. My work is partially supported by the ERC grant in the framework of the European Union ``Horizon 2020'' program (QUASIFT, grant №~677368).

\vspace{0.7cm}

\end{document}